\documentstyle[12pt]{article}
\newcommand{\gsim}
{\raisebox{-0.7ex}{$\;\stackrel{\textstyle >}{\textstyle\sim}\;$}}

\newcommand{\lsim}
{\raisebox{-0.7ex}{$\;\stackrel{\textstyle <}{\textstyle\sim}\;$}}

\begin{document}
\begin{center}

{\bf  
POWER TAILS OF ELECTRIC FIELD DISTRIBUTION FUNCTION  IN 2D 
METAL-INSULATOR COMPOSITES
} \vspace{0.5cm}

{ E.M.Baskin, M.V.Entin}

{\small \it Institute of Semiconductor Physics, Russian Academy of
Sciences, Siberian Branch, Prosp. Lavrentyeva, 13, Novosibirsk
630090, RUSSIA.  }

\end{center}

{\small
The 2D "Swiss-cheese" model of conducting media with round
insulator inclusions is studied in the 2nd order of
inclusion concentration and near the percolation threshold. The
electric field distribution function is found to have  power
asymptotics for fields much exceeding the average field,
independently on the vicinity to the threshold, due to finite
probability of arbitrary proximity of inclusions. The strong field 
in the narrow necks between inclusions results in the induced 
persistent anisotropy of the system. The critical index for noise 
density is found, determined by the asymptotics of electric field 
distribution function.  }\\ \vspace{0.3cm}

\subsubsection*{Introduction}
The distribution function of electric field is
an important characteristics of a conducting mixture.  It determines
such properties as noise, electric breakdown and nonlinear dielectric
permeabilities.
Its long-range tails are usually considered as a result of the electric 
field amplification due to the vicinity to the percolation threshold 
\cite{1}, \cite{2}, \cite{3}.

The purpose of the present study is to find the distribution function
for the electric field $\bf F$ in the conducting phase. The 2D model of
conducting media containing randomly distributed round voids with
density $n$ and radius $a$ is considered. This system is interesting
both as a real object of nanoelectronics \cite{4} and as an exact
solvable model.  
Without inclusions the electric field is uniform. The field around 
a solitary smooth void is limited, so it does not produce the field, 
much exceeding the average field. The strong field appears in 
the 2nd order of void concentration only.

We shall study both 2nd order of concentration and the field 
distribution at the percolation threshold.

\subsubsection*{Potential distribution around two
holes.}
The distribution of field in a conducting media is described by the
solution of the continuity equation ${\bf \nabla j}=0$ for current
${\bf j}=\sigma {\bf F}=\sigma {\bf \nabla}\varphi$, where $\sigma$
is the conductivity of the conductor.  The boundary conditions on
the void boundaries  require vanishing of the normal component of 
current density $\sigma {\bf n}\nabla \varphi$ (${\bf n}$ is the 
normal to the circle). The electrostatic problem of field 
distribution is two-dimensional.

Since the asymptotics of the field distribution is determined by  
narrow necks  between two holes, we turn our attention on the
media , containing two round holes only (Fig. 1).
We assume that the $x$ axis goes through circles centers. The 
conformal mapping of system of Fig.1 to the band $1/R_0<Re w<R_0$ is 
given by the function \cite{Lavren}

\begin{equation}
w=\frac{z+b}{z-b},
\end{equation}

where $z=x+iy$, $2\delta$ is the distance between the holes with 
radii $a$, $$ b=\sqrt{2a\delta +\delta^2},~~~~~~~~ R_0= 
\frac{\sqrt{2a+\delta}+\sqrt{\delta}}{\sqrt{2a+\delta}-\sqrt{\delta}}
$$
We should solve the problem of current flow with conditions 
$\varphi\sim \mbox{~Im~}(F_0z)$ for $z\to \infty$, where $F_0$ is the 
average electric field.  The corresponding potential in $w$-plane is 
the dipole potential with asymptotic behavior $ \varphi\sim 
\mbox{~Im~}(\frac{2F_0b}{w-1}), $ for $w \to 0$.

The dipole potential  in the ring  between two circles is 
determined by the sum of its images in circles:  
\begin{equation} 
\varphi=2F_0b~~\mbox{ 
Im}~\Bigl( \frac{1}{w-1}+\sum_{n=1}^{\infty}\frac{1}{w R_0^{2n}-1}-
\sum_{n=1}^{\infty}\frac{w}{R_0^{2n}-w} \Bigr)
\end{equation}
The strong field results from the close approaching of two circles,
$\delta\ll a$. In this limit (\theequation) transforms:
\begin{equation}
\varphi=\pi F_0a~ \mbox{Im}\bigl(\cot{\frac{\pi a}{z}}\bigr)
\end{equation}
The last extreme limit means the absence of percolation between
 contacting circles if $\delta \to 0$. This leads to the gap of
 potential between up and down sides of the circles near the points of
 tangency:
 \begin{equation}
 U=2\pi F_0 a
 \end{equation}

If the width of constriction is small but finite, the potential gap
 is distributed along the length of constriction $\sqrt{2a\delta}$.
 The field in the vicinity of constriction is
 \begin{equation}
\label{7}
 F_y=\frac{U}{\pi}\sqrt{\frac{2\delta}{a}}\frac{1}{2\delta+y^2/a}
 \end{equation}
The maximal value of field  $ F_{max}(\delta) = F_0
 \sqrt{2a/\delta}$ diverges with $\delta\to 0$. Another 
component of field, $F_x$, remains finite if $\delta\to 0$ and 
is insufficient below.

\subsubsection*{Breakdown-induced training.}

The divergency of field and current density can lead to overheating
and melting of narrow necks. If we assume that breakdown is determined
by the critical density of heat emission $Q_c$, all places where
$\sigma F^2>Q_c$ will be overheated. 
It means that some number of overheated places occurs 
in any weak field $F_0$ independently on the  field strength. So  the 
application of field irreversibly change the electrical properties of 
system. Moreover, the direction of field tends to the residual 
anisotropy of a system. 

We shall consider this effect in the framework of swiss-cheese 
system with rare, randomly distributed round voids.  The 
residual change of linear conductivity in a low field originates from 
the breakdown of narrowest necks. The criterium of breakdown is $ 
 F_{max}(\delta)\geq F_c = 
\sqrt{Q_c/\sigma}$.

The contribution of necks between pairs of voids which have 
x-orientation and widths between $\delta$ and $\delta+d\delta$ to the 
average conductivity is
\begin{eqnarray} 
\Delta \sigma_{yy}=16 \pi \sqrt{2} (na^2)^2 ~\sigma ~
\sqrt{\frac{\delta}{a}}\frac{d\delta}{a},\\ 
\nonumber
\Delta \sigma_{xx}=0.  
\end{eqnarray} 
One should average (\theequation) over $\delta$ and 
the direction of pair of voids. The value $\delta$  is limited by the 
maximum $\delta_{max}$ determined by the situation when the field in 
the neck exceeds the breakdown field $F_{c}=F_0 
\sqrt{\frac{\delta_{max}}{a}} \cos \theta $, where $\theta$ is the 
angle between the external field and the direction  of  axis, 
connecting centers of circles.  The resulting change of conductivity 
tensor is \begin{eqnarray} \Delta \sigma_{xx}&=&- 
\frac{128\sqrt{2}}{45}(na^2)^2 \Bigl(\frac{|{\bf 
F}_0|}{F_{c}}\Bigr)^3~\sigma ~\\ \Delta \sigma_{yy}&=&4 \Delta \sigma_{xx}.  
\end{eqnarray} The direction $y$ is determined along the training 
field. 

The change of longitudinal conductivity appears to be 4 times 
higher than transversal. The equation (\theequation) shows the 
non-analytical dependence on the field. It should be noted that the 
training, being a nonlinear effect in applied field, nevertheless 
means the change of linear conductivity. 

 \subsubsection*{Distribution function of field for pairs of circles.}
The asymptotics  of distribution function for high field is determined
by rare approaching of circles, when the distance $\delta$ between
them is much less than the radius and mean distance $(\pi n)^{-1/2}$.

The probability of field to have value $F$, $P_{\delta,U} (F)$ for 
the fixed distance $\delta$  and potential gap $U$ is determined by 
area where $F$ is between $F$ and $F+dF$.  By means of (\ref{7}) we 
find:

 \begin{equation}
\label{200}
 P_{\delta,U} (F)=
(\frac{2\delta}{\pi^2})^{3/4}
\frac{1}{a^{1/4}}
\frac{U^2}{F^{5/2}(U-F\pi\sqrt{2a\delta})^{1/2}}
\end{equation}
This result should be integrated by $\delta$ with the
density of pairs $ 4 \pi n^2 a~~ d\delta$, distance of which is 
within the range $2\delta$ and $2(\delta+d\delta)$ .  The result of 
averaging on $\delta$ and the direction of axis $y$ is 
\begin{equation} 
\label{10} P(F)= \frac{128}{3\pi }(\pi 
na^2)^2\frac{F_0^5}{F^6} 
\end{equation}

The equation (\theequation)  gives
power  long-range tail of the electric field distribution function
$P(F) \sim F^{-6}$.

The long-range distribution tail gives the divergency of high order power
of the local field, starting from 6th power and does not affect 
the density of $1/f$ noise, determined by 4th power of ${\bf F}$.

The existence of power tail results from the
microgeometry and is not connected with the vicinity to the 
percolation threshold. Thus such power tails are universal property 
of 2D system both far and near threshold.  They result from the 
possibility for 2D system to have a finite potential gap $U$ on 
infinitely small distance, connected with the necessity of current to 
flow around tangent circles on the finite distance to equilibrate 
the potential gap. On the other  hand, the shortest path between two 
infinitely close points along the conducting phase in the 3D case is 
always small and the field near close pair of inclusions is 
limited.  Hence, no long-range tail exists in 3D system of 
non-intersecting spheres and only exponential tail $P(F)$ is possible 
in the intersecting inclusions 2D or 3D models determined by rare 
large coupling of inclusions.

\subsubsection*{Percolational situation.}
The power-law tails due to formula (\ref{10}) are determined by very
small part of total current while most part  of current flows around
both disks.  Hence in the percolational limit $na^2\sim 1$
the contribution retains,  described by random necks inside the 
conducting phase. Formula (\ref{10}) may be applied to this case 
also (except for numerical coefficient) if one changes $na^2\to 1$.

The key bonds tend to
additional possibility for local field amplification near the
threshold.

The distribution 
function of voltages in  lattice models was studied on the metal 
side of percolation threshold  \cite{1,tail}. It was found that this 
function has a long-range tail, $P(U)\sim 
exp((\mbox{ln}(U/U_0))^2/C)$, which is much weaker than any 
power-like one. The distribution function in insulator phase, found 
below, exhibits power behavior.  In the Swiss-cheese model narrow 
necks, insufficient for current flow may be considered as almost 
insulating phase, so these tails are reproducing themselves in the 
contribution from narrow necks to metal phase distribution function 
long tail.

To estimate the insulator phase distribution function tail we shall
use the evident formula for energy:
\begin{equation}
\label{energy}
\frac{\langle\varepsilon {\bf F}^2\rangle}{8\pi}=
          \varepsilon_{e}  \frac{{\langle \bf F\rangle}^2}{8\pi}
\end{equation}
where $\varepsilon$ is the permittivity of the insulator. Just below
percolation threshold the effective permittivity $\varepsilon_{e} $ 
diverges like $\tau^{-q}$, producing divergency of $\langle {\bf 
F}^2\rangle$ if the system comes to the percolation threshold $\tau 
=0$, while the mean field $\langle \bf F\rangle$ is limited.  These 
two facts do not contradict if the distribution function behaves like 
$1/F^{3-r}$, where $0<r<1$.

Near the percolation threshold one should expect the scaling behavior 
of a distribution function, both above and below the threshold. 
The characteristic scale for  voltage on single bond $U=Fa$ is 
determined by the maximal voltage $U_m$ in a percolation cell. 
This voltage  can be estimated by assumption, that full voltage across 
the cell with size $L_c$ is applied to one disconnected bond: 
$U_m\sim F_0L_c\sim F_0 a\tau^{-\nu}$.  Hence the distribution 
function for voltages is

\begin{equation}
\label{100}
p(U)=\frac{1}{F_0 a}(F_0a/U)^{3-r}f(\frac{U}{F_0a}\tau^\nu)
\end{equation}

The expression for exponent $r=q/\nu$  is determined by equation 
(\ref{energy}). According to known values of critical exponents in 
2D case $r=0.9$.  This result  is consistent with numerical result of 
\cite{Chakr} but  slightly deviates from estimation based on the 
assumption, that asymptotics $p(U)$ is determined by the number of 
single-disconnected bond per a percolation cell \cite{5}  $N_{sd}\sim 
\tau^{-1}$:  $p(U_m)\sim \frac{N_{sd}}{U_m L_c^2}\sim 
\tau^{3\nu-1}$. Like so called nodes-links-blobs model  we shall 
disregard small deviations from 1 of critical  exponents  for 
effective conductivity $\sigma_{e}\sim \tau^t$ and effective 
permittivity $\varepsilon_e\sim \tau^{-q}$.  
 
The far tail for electric field distribution function in the 
Swiss-cheese model on the metal side of percolation transition is 
determined by narrow necks on the infinite cluster. The strongest 
voltage  is expected from a narrow neck on a branch of infinite 
cluster clumped to the backbone  in two points with distance $L_c$ 
between them. The width of the neck is supposed so small, that it 
resistance exceeds the backbone resistance and current through it is 
negligible.  Hence the typical width is   limited by $\delta\lsim 
\tau^{2t}$.  Such necks may be considered as insulating for the 
problem of current flow.  For  the neck with the potential gap has 
order of magnitude $U=F_0L_c$. Integrating the distribution function 
of field  (\ref{200})   together with the distribution function of 
voltages  we find the distribution function of field in 
the vicinity of percolation threshold
 
\begin{eqnarray} \label{201}
P(F)\sim 
\tau^{-(t+3\nu)} \frac{F_0^{5}}{F^{6}}~~~~~\mbox{for}~~ F \gsim F_0 
\tau^{-(t+\nu)}\\
\label{202}
P(F)\sim 
\tau^{(5t+\nu)/2} \frac{F_0^{3/2}}{F^{5/2}} ~~~~~\mbox{for}~~ F \lsim 
F_0 \tau^{-(t+\nu)}.  \end{eqnarray}
Comparison of formulae (\ref{201}), (\ref{202}) and (\ref{10}) shows 
that the validity of (\ref{202}) is limited by an inequality $ 
F_0 \tau^{-(5t+\nu)/7}\lsim F \lsim F_0 \tau^{-(t+\nu)}$ where the 
contribution of single-disconnected bonds prevails. For lower fields 
$F\lsim F_0 \tau^{-(5t+\nu)/7}$ the asymptotics of distribution 
function is given  by the contribution of narrow necks inside metal 
phase (\ref{10}).

The amplification of field distribution function tails near the 
percolation threshold leads to enhancement of current noise. The 
integral noise in disordered media is determined by mean 4th power of 
field, according to \cite{Ra}:
\begin{equation}
C^e=\frac{\langle\sigma^2F^4\rangle}{(\langle 
{\bf F}\rangle\langle\sigma {\bf F}\rangle)^2}. \end{equation}
The substitution of (\ref{201}, \ref{202}) to (\theequation) gives 
the critical exponents for integral noise:
\begin{equation}
C^e\sim \tau^{-2(t+\nu)}.
\end{equation}

\subsubsection*{Concluding remarks.}
Let us summarize the obtained results.

We found the field distribution around pair of circular 
insulating inclusions into the conducting phase.  It was used to find 
the second virial correction on the density of inclusions to the 
distribution function of field.  The distribution function was found 
to have the power-like long range tail, determined by close 
locations of inclusions. The same configurations result in 
the training of medium, that is, an irreversible modification 
of material by a weak electric field, which 
destroys the narrow necks and produces the persistent anisotropy. 

It was demonstrated, that the tail of distribution function is 
enhanced near the percolation threshold. 
The field distribution function was employed to find the current 
noise in the system. It was found that it is amplified of the 
near the percolaton threshold. The  main sources of this 
amplification are the narrow necks belonging to the percolation 
cluster backbone.

The results of the present article have close tights to the 
system dimensionality.  They are determined by the possibility 
of existence  of a finite potential gap between two close points in 
2D, in contradiction with 3D system. This conclusion is independent 
on the model of inclusions we considered.

 The work was partially supported by Russian Foundation 
for Basic Researches (Grants 950204432 and 960219353) and 
Folkswagen-Stiftung.   \newpage 

\unitlength=2pt 
\begin{picture}(100,100)
\put(30,50){\vector(1,0){70}}
\put(55,20){\vector(0,1){65}}
\put(60,50){\line(0,-1){30}}
\put(47,24){\vector(1,0){8}}
\put(68,24){\vector(-1,0){8}}

\thicklines
\put(10,5){{\bf Figure} 1. System with 2 round holes. }
\large
\put(41,52){ \it a}
\put(56,80){$y$}
\put(97,52){$x$}
\put(56,22){$\delta$}
\put(85,67){${\bf F}_0$}
\put(40,50){\circle{1}}
\put(40,50){\vector(1,2){4.01}}
\put(95,60){\vector(0,1){20}}
\put(40,50){\circle{40}}
\put(70,50){\circle{40}}
\end{picture}

\end{document}